\documentclass[prb,twocolumn,superscriptaddress,floats,citeautoscript,showpacs]{revtex4}

\usepackage{graphicx}
\usepackage{amsmath}
\usepackage{amssymb}
\usepackage{times}
\usepackage{colordvi}
\usepackage{bm}

\newcommand* {\vek}[1]{{\ensuremath{\bm{\mathrm{#1}}}}}
\newcommand* {\ee}{\ensuremath{\mathrm{e}}}
\newcommand{\ket}[1]{\left |  #1 \right \rangle}

\graphicspath{{.}{./EPS/}}

\begin{document}

\title{Tunable Aharonov-Anandan Phase in Transport Through Mesoscopic Hole
Rings}

\author{M. Pletyukhov}
\affiliation{Institut f\"ur Theoretische Festk\"orperphysik and DFG Center for
Functional Nanostructures, Universit\"at Karlsruhe, D-76128 Karlsruhe, Germany}
\affiliation{Institut f\"ur Theoretische Physik A, RWTH Aachen, D-52056 Aachen,
Germany}
\altaffiliation[Present address.]{}

\author{U. Z\"ulicke}
\affiliation{Institut f\"ur Theoretische Festk\"orperphysik and DFG Center for
Functional Nanostructures, Universit\"at Karlsruhe, D-76128 Karlsruhe, Germany}
\affiliation{Institute of Fundamental Sciences and MacDiarmid Institute for Advanced
Materials and Nanotechnology, Massey University, Private Bag 11~222, Palmerston
North, New Zealand}
\altaffiliation[Permanent address.]{}

\date{\today}

\begin{abstract}

We present a theoretical study of spin-3/2 hole transport through mesoscopic rings,
based on the spherical Luttinger model. The quasi-one-dimensional ring is created in
a symmetric two-dimensional quantum well by a singular-oscillator potential for the
radial in-plane coordinate. The quantum-interference contribution to the two-terminal
ring conductance exhibits an energy-dependent Aharonov-Anandan phase, even
though Rashba and Dresselhaus spin splittings are absent. Instead,
confinement-induced heavy-hole--light-hole mixing is found to be the origin of this
phase, which has ramifications for magneto-transport measurements in gated hole
rings.

\end{abstract}


\maketitle

{\it Introduction.}\/
Geometric quantum phases continue to be the subject of great interest because they 
rather elegantly elucidate quite complex fundamental microscopic
properties~\cite{wilczekBook}. The well-known Aharonov-Bohm~\cite{ahabohm} and
Aharonov-Casher~\cite{ahacash} effects are pertinent examples, as is the Berry
phase~\cite{berry} acquired by a quantum system during adiabatic cyclical variation
of an external parameter. The more general concept of the Aharonov-Anandan
phase~\cite{ahaanan} arising in the cyclical evolution of a quantum state was later
shown to subsume the above-mentioned three phenomena as special cases. Recent
progress in our understanding of geometric phases has been spurred by numerous
experimental and theoretical studies~\cite{anan:ajp:97}. 

Modern nanofabrication techniques have made it possible to study
quantum interference, and thus geometric quantum phases, in mesoscopic
electronic devices~\cite{mesotrans}. The first theoretical
suggestions~\cite{gefen:prl:84} and experimental realizations~\cite{aboscmetal} of
electronic quantum interference devices were Aharonov-Bohm interferometers.
Subsequent theoretical studies predicted electronic signatures of Berry
phases~\cite{ady:prl:92} and the Aharonov-Casher
effect~\cite{harsh:prl:92,nitta:apl:99,jairo:prb:07}. Inspired by possible applications in
the burgeoning field of spintronics~\cite{zutic:rmp:04}, recent experimental efforts
have been devoted to observing tunable spin-related  geometric phases in
semiconductor rings subject to strong spin-orbit
coupling~\cite{nitta:prl:06,mole:prl:06}. In several of these
experiments~\cite{mole:prl:06}, charge carriers moving through the ring structure are
characterized by an intrinsic (spin) angular momentum equal to 3/2. This is due to
the fact that states in the uppermost valence band of typical semiconductors originate
from p-like atomic orbitals of the material's constituent elements~\cite{rolandbook}. 
The same is true for conduction-band states in HgCdTe quantum wells because of a
band inversion. Here we provide a careful study of the complexities
arising from the spin-3/2 character of charge carriers confined in a mesoscopic ring.
We identify a nontrivial part of the Aharonov-Anandan (AA) phase that enters the
interference contribution to the two-terminal ring conductance. The energy
dependence of this geometric phase will result in a continuous shift of
magneto-conductance (Aharonov-Bohm) oscillations as a function of carrier density
in the ring, e.g., when a gate voltage is applied. We trace the origin of the
anomalous AA phase to a confinement-induced coupling between heavy-hole (HH)
and light-hole (LH) states~\cite{HHlhComm} that is ever-present and unrelated to
(Rashba or Dresselhaus) spin splitting due to (structural or bulk) inversion
asymmetry~\cite{rolandbook}. Our analysis provides a framework for interpreting
numerical results~\cite{mingwei:pla:06} and complements previous analytical
calculations~\cite{jairo:prb:07} where HH-LH mixing was neglected.

Below we describe our theoretical model for hole rings. Readers not interested in
mathematical details could skip to the end of this part where results for the lowest
ring subbands are presented. We then analyse the emerging energy-dependent
AA phase and, in our concluding discussion, address implications for experiments.

{\it Theoretical model for a mesoscopic hole ring.}\/
We use the Luttinger model~\cite{luttham2} in spherical
approximation~\cite{lip:prl:70} to describe electronic states in the upper-most bulk
valence band. In atomic units where $\hbar = m_0 = 1$, it reads
\begin{widetext}
\begin{eqnarray}
{\mathcal H}_{\text{s}} &=& \left( \frac{\gamma_1}{2} + \frac{\gamma_{\text{s}}}{2}
\left[ {\hat J}_z^2 - \frac{5}{4} \right] \right) {\hat k}_\perp^2 - \frac{\gamma_{\text{s}}}
{2}  \left( {\hat k}_-^2 {\hat J}_+^2 + {\hat k}_+^2 {\hat J}_-^2 \right) + \left( \frac
{\gamma_1}{2} - \gamma_{\text{s}} \left[ {\hat J}_z^2 - \frac{5}{4} \right] \right)
{\hat k}_z^2 \nonumber \\
&& \hspace{7cm} - \sqrt{2} \gamma_{\text{s}} \left( \left\{ {\hat k}_z , {\hat k_-} \right\}
\left\{ {\hat J}_z , {\hat J_+} \right\} + \left\{ {\hat k}_z , {\hat k_+} \right\} \left\{
{\hat J}_z , {\hat J_-} \right\} \right) \quad .
\end{eqnarray}
\end{widetext}
$m_0$ denotes the vacuum electron mass, $\gamma_{\text{s}}=(3 \gamma_3
+ 2 \gamma_2)/5$ in terms of Luttinger parameters~\cite{luttham2}, $\vek{\hat k}$
and $\vek{\hat J}$ are vector operators of kinetic linear and spin-3/2 angular
momentum, respectively, and we used the abbreviations $\vek{\hat k_\perp} = (\hat
k_x, \hat k_y)$,  ${\hat k}_\pm = {\hat k}_x\pm i {\hat k}_y$, and ${\hat J}_\pm = 
({\hat J}_x \pm i {\hat J}_y )/\sqrt{2}$.
The symbol $\{ A , B \}$ stands for the anticommutator $(AB + BA)/2$.
Introducing a quantum-well confinement in the 
growth ($z$) direction, two-dimensional (2D) subbands are formed. Here we will
focus on the situation where only the lowest 2D quantum-well bound state matters.
To be specific, we assume a symmetric hard-wall confinement with 2D quantum-well
width $d$ and simply replace operators ${\hat k}_z^2$ and ${\hat k}_z$ by their
respective expectation values $\pi^2/d^2$ and $0$~\cite{KzComm}. Using polar
coordinates for in-plane motion, we obtain ${\mathcal H}_{\text{s}}^{\text{(2D)}} =
{\mathcal H}_{\text{sb}} + {\mathcal H}_{\perp}^{\text{(2D)}}$, where
\begin{subequations}
\begin{equation}\label{parHam}
{\mathcal H}_{\text{sb}} = \left( \frac{\gamma_1}{2} - \gamma_{\text{s}} \left[
{\hat J}_z^2 - \frac{5}{4} \right] \right) \left( \frac{\pi}{d} \right)^2
\end{equation}
arises from the quantised motion in $z$ direction, and the in-plane motion of holes is
governed by the part
\begin{widetext}
\begin{equation}\label{perpHam}
{\mathcal H}_{\perp}^{\text{(2D)}} = \left( \frac{\gamma_1}{2} +
\frac{\gamma_{\text{s}}}{2} \left[ {\hat J}_z^2 - \frac{5}{4} \right] \right) \left\{ -
\partial_r^2 -\frac{\partial_r}{r} + \frac{{\hat L}_z^2}{r^2} \right\} -
\frac{\gamma_{\text{s}}}{2}  \left( {\hat J}_+^2 \left[ -i L_- \left\{ \partial_r +
\frac{{\hat L}_z}{r} \right\} \right]^2  +  {\hat J}_-^2 \left[ -i L_+ \left\{ \partial_r -
\frac{{\hat L}_z}{r} \right\} \right]^2 \right) .
\end{equation}
\end{widetext}
\end{subequations}
${\hat L}_z = -i \partial_\varphi$ is the in-plane orbital angular momentum, and
$L_\pm = \ee^{\pm i \varphi}$. The Hamiltonian (\ref{perpHam}) commutes with $\hat
M_z = \hat L_z + \hat J_z$; hence its eigenstates can be labelled by those of $\hat
M_z$. To enable further analytic progress, we eliminate the $\varphi$ dependence in
off-diagonal matrix elements by the transformation $\tilde{\mathcal H} = \ee^{i {\hat
J}_z \varphi} \, {\mathcal H} \,\ee^{-i {\hat J}_z\varphi}$. Due to space limitations, the
straightforwardly obtained expression for the transformed Hamiltonian is omitted
here.

The quasi-onedimensional in-plane ring confinement is modeled by the
singular-oscillator potential~\cite{tan:sst:96}
\begin{equation}\label{eq:SingOsc}
V_\perp(r) = \frac{\omega^2}{2} \left( r - \frac{R^2}{r} \right)^2 \quad ,
\end{equation}
which was employed before to study mesoscopic {\em electron\/}
rings~\cite{tan:prb:99}. In the following, the energy scale $E_0 = \pi^2 \gamma_1
\hbar^2 / (2 m_0 d^2)$ associated with the 2D quantum-well confinement will serve
as our energy unit. We also introduce the parameter $\bar\gamma=
\gamma_{\text{s}}/\gamma_1$ that measures the strength of spin-orbit coupling in
the valence band, the length scale $\ell_\omega = \sqrt{\sqrt{\gamma_1} \hbar/(m_0
\omega)}$ associated with the ring confinement, $\lambda_R = \left( R /\ell_\omega
\right)^2$ and $\lambda_d = \left(2 d / [ \pi \ell_\omega]\right)^2$ representing ring
radius and quantum-well width in units of the effective ring width, and the operators
$\hat\Gamma = 1 + \bar\gamma \left[ {\hat J}_z^2 - (5/4)\right]$ and $\hat\varrho = r/
(\ell_\omega \hat\Gamma^{1/4})$ that quantify a HH-LH splitting. With these 
conventions, the Hamiltonian of a mesoscopic hole ring is $H \equiv (\tilde{\mathcal 
H}_{\text{s}}^{\text{(2D)}} + V_\perp)/E_0 = H_{\text{qw}} + H_{\text{rg}}$, where
\begin{subequations}
\begin{equation}
H_{\text{qw}} = 1 - 2 \bar\gamma \left[ {\hat J}_z^2 - \frac{5}{4} \right]
\end{equation}
arises from the HH-LH splitting in the quantum-well bound state, and the in-plane
motion is governed by
\begin{widetext}
\begin{equation} \label{ringHam}
H_{\text{rg}} = \frac{\lambda_d}{4} \, {\hat\Gamma}^{\frac{1}{2}} \left( -\partial_{\hat
\varrho}^2 -\frac{\partial_{\hat\varrho}}{\hat\varrho} + \frac{{\hat {\tilde L}}_z^2}{\hat
\varrho^2} + \left[ \hat\varrho - \frac{\lambda_R \, {\hat\Gamma}^{-\frac{1}{2}}}{\hat
\varrho} \right]^2 \right) + \frac{\lambda_d}{4} \, \bar\gamma \left\{ \left( {\hat J}_+ \,
{\hat \Gamma}^{-\frac{1}{4}} \left[ \partial_{\hat\varrho} + \frac{{\hat {\tilde L}}_z}{\hat
\varrho} \right] \right)^2 + \left( {\hat J}_- \, {\hat \Gamma}^{-\frac{1}{4}} \left[
\partial_{\hat\varrho} - \frac{{\hat {\tilde L}}_z}{\hat\varrho} \right] \right)^2 \right\} \, .
\end{equation}
\end{widetext}
\end{subequations}

\begin{figure}[b]
\includegraphics[width=2.3in]{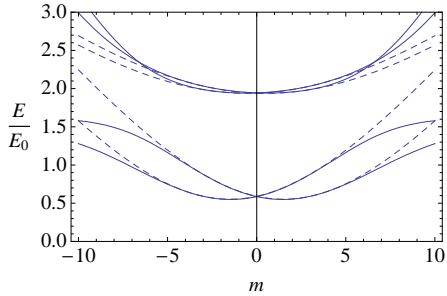}
\caption{\label{fig:disp}
Lowest hole-ring subbands (solid curves) arising from the in-plane ring-potential
bound-state level with $n=0$, calculated for $\bar\gamma=0.37$, $\lambda_d=0.5$, 
and $\lambda_R=10$. The dashed curves are obtained when HH-LH \textit{splitting}
is included but HH-LH \textit{mixing} is neglected.}
\end{figure}

Equation~(\ref{ringHam}) suggests the wave-function \textit{ansatz}
\begin{subequations}\label{ringAnsatz}
\begin{equation}
\psi_{n,m}(r, \varphi) = \ee^{i m \varphi}  \, \left( \begin{array}{c} a_{n,m}\,
\psi_{n,m}^{(3/2)}(r)  \\ b_{n,m}\, \psi_{n,m}^{(1/2)}(r)   \\ c_{n,m}\, \psi_{n,m}^{(-1/2)}
(r) \\ d_{n,m}\, \psi_{n,m}^{(-3/2)}(r) \end{array} \right) ,
\end{equation}
with the four spinor amplitudes given by
\begin{equation}
\psi_{n,m}^{(j)}(r) = \frac{{\mathcal N}_{n,m}^{(j)}}{\ell_\omega [1 + 2(|j|-1) \bar
\gamma ]^{1/4}} \,\, \ee^{-\frac{\varrho_{j}^2}{2}} \, \varrho_{j}^{\alpha_{m}^{(j)}} \,
{\mathrm L}_n^{(\alpha_m^{(j)})} \!\! \left( \varrho_{j}^2 \right) \, .
\end{equation}
\end{subequations}
${\mathrm L}_n^{(\alpha)}(x)$ is an associated Laguerre polynomial,
${\mathcal N}_{n,m}^{(j)} = \left[2 \Gamma(n + 1)/\Gamma(n + \alpha_m^{(j)} + 1)
\right]^{1/2}$ with $\Gamma(x)$ denoting the Euler Gamma function, $\varrho_{j} = r
/ \left(\ell_\omega [1 + 2(|j|-1) \bar \gamma]^{1/4} \right)$, and $\alpha_m^{(j)}=\sqrt
{(m - j)^2 + \lambda_R^2 /(1 + 2(|j|-1) \bar \gamma)}$. $m$ is the eigenvalue of
$\hat M_z$ and $n$ the oscillator-level index. \textit{Ansatz\/} (\ref{ringAnsatz})
diagonalises the first term of the Hamiltonian~(\ref{ringHam}),
\begin{equation}
H_{\text{rg}}^{(1)} = \lambda_d \left( \left[ n + \frac{1}{2} \right] {\hat\Gamma}^{\frac
{1}{2}} + \frac{\sqrt{\hat\Gamma \left( m - \hat J_z \right)^2 + \lambda_R^2} -
\lambda_R}{2} \right) \! .
\end{equation}
However, the second term of $H_{\text{rg}}$ in Eq.~(\ref{ringHam}) is off-diagonal in 
spin-3/2 space, coupling HH and LH amplitudes within subspaces spanned by $\hat
J_z$ eigenstates with eigenvalues $\{\pm 3/2, \mp 1/2\}$, respectively. In addition,
this term has both diagonal and  off-diagonal matrix elements in the $\psi_{n,m}$
representation, i.e., it couples states with different $n$. We omit the lengthy
analytical expressions for associated matrix elements. It turns out that HH-LH mixing
between states having their quantum number $n$ differ by 0, 1, and 2 are most
relevant. In the following, we focus on the lowest oscillator level ($n=0$) and include
only its intra-level HH-LH mixing. Neglecting the subtle difference between
$\varrho_j$ and $\varrho_{j\pm 2}$, setting both equal to $r/\ell_\omega$, and 
replacing $\alpha_m^{(j)} \equiv \lambda_R$ yields the corresponding matrix 
element~\cite{ApproxComm}
\begin{widetext}
\begin{equation}
\left( H_{\text{rg}}^{(2)} \right)_{00} = - \frac{\lambda_d \, \bar\gamma}{4} \left\{
\hat J_+^2 \left[ 1 - \frac{( m - \hat J_z - 2)( m - \hat J_z )+2}{\lambda_R} \right] +
\hat J_-^2 \left[ 1 - \frac{( m - \hat J_z + 2)( m - \hat J_z )+2}{\lambda_R} \right]
\right\} . 
\end{equation}
\end{widetext}
Diagonalising the Hamiltonian $H_{\text{qw}} + H_{\text{rg}}^{(1)} + \big(
H_{\text{rg}}^{(2)} \big)_{0 0}$ yields the lowest hole-ring subband dispersions. The 
result is shown in Fig.~\ref{fig:disp} for a set of realistic parameters. For comparison, 
we also plot dispersions obtained when HH-LH {\em mixing\/} is neglected but
HH-LH {\em splitting\/} taken into account, i.e., when only $H_{\text{qw}} +
H_{\text{rg}}^{(1)}$ is considered.

{\it Two-terminal transport and AA phase.}\/ 
\begin{figure}[b]
\includegraphics[width=1.7in]{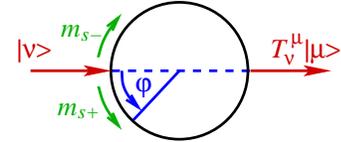}
\caption{\label{fig:ringScatt}
(Color online) Two-terminal transport. Holes in scattering states $\ket{\nu}$ are
injected via a lead attached at $\varphi=0$. Eigenmodes with angular momenta
$m_{s\sigma}$ provide propagation channels in the ring. Upon reaching the point
$\varphi=\pi$, interference and coupling into outgoing scattering states $\ket{\mu}$
occurs.}
\end{figure}
At fixed energy $E$ between the HH-like and LH-like subband bottoms shown in
Fig.~\ref{fig:disp}, four propagating channels exist, having different angular momenta
$m_{s \sigma}$. Here $s=+,-$ labels the two dispersion curves ($s=\pm$
corresponding to the subspace spanned by $\hat J_z$-projection eigenstates with
eigenvalues $\{\pm 3/2, \mp 1/2\}$, respectively), and $\sigma=+,-$ distinguishes
opposite propagation directions. The condition for finding these angular momenta is
$E_s(m_{s \sigma}) =E$. We now consider the following scenario, illustrated in
Fig.~\ref{fig:ringScatt}. A lead attached at $\varphi =0$ is assumed to inject holes in
a set of orthogonal initial states $\ket{\nu}$. Here the ket refers to a normalised
spin-3/2 spinor without any dependence on spatial coordinates~\cite{ProfComm}.
Ring eigenstates taken at $\varphi=0$ and with spatial profile neglected are
given by $\chi_{s \sigma}=(a_{0, m_{s \sigma}}, b_{0, m_{s \sigma}}, c_{0, m_{s
\sigma}}, d_{0, m_{s \sigma}})$, and the injected states can be  written as a
superposition
$\ket{\text{in}}_\nu = \sum_{s \sigma} \, \xi^{(\nu)}_{s \sigma} \, \chi_{s \sigma}$.
Each spinor amplitude of a ring eigenstate with quantum numbers $s,\sigma$
acquires a phase during propagation around a half-ring arm that is determined by its
associated eigenvalue of the operator $\sigma \pi ({\hat M}_z - {\hat J}_z)$. The
resulting state at $\varphi = \pi$ will be
$\ket{\text{out}}_\nu = \sum_{s \sigma} \, \xi^{(\nu)}_{s \sigma} \, \chi_{s \sigma} \,
\exp\{{\sigma i \pi ( m_{s \sigma} - \frac{3 s}{2} )\}}$.
Assuming another lead being attached at $\varphi=\pi$, we find the transmission
probabilities from incoming-lead channel $\nu$ to outgoing-lead channel $\mu$.
Including the effect of a finite magnetic flux $\phi$ threading the area bounded by the 
ring, it reads
\begin{equation} \label{eq:transm}
T_\nu^\mu = \left| \sum_{s \sigma}\, \xi^{(\mu) \ast}_{s \sigma}\, \xi^{(\nu)}_{s \sigma}
\, \ee^{\sigma i \pi \left( m_{s \sigma} + \frac{\phi}{\phi_0} - \frac{3 s}{2} \right)}
\right|^2 \quad . 
\end{equation}
In Eq.~(\ref{eq:transm}), $m_{s\sigma}\equiv -m_{-s,-\sigma}$ are the
angular-momentum eigenvalues found for $\phi=0$. The linear two-terminal ring
conductance is given by
$G_{\text{rg}} = \frac{e^2}{2\pi\hbar} \sum_{\nu, \mu} T_\nu^\mu$.

\begin{figure}[t]
\includegraphics[width=2.2in]{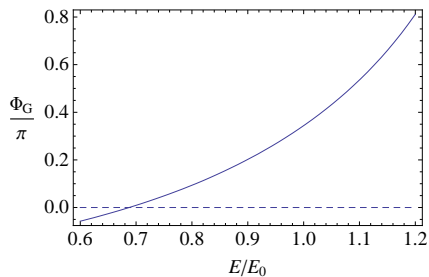}
\caption{\label{fig:Berry}
Solid curve: Anomalous component $\Phi_{\text{G}}$ of the AA phase appearing in
the two-terminal hole-ring transmission. Parameters used in the calculation are the
same as in Fig~\ref{fig:disp}. Dashed curve: Corresponding result obtained when
HH-LH mixing is neglected.}
\end{figure}

Inspection of Eq.~(\ref{eq:transm}) reveals the well-known signatures of the
Aharonov-Bohm effect~\cite{gefen:prl:84}, arising from the interference 
of counter-propagating modes with conserved quantum number $s$. A phase 
difference $\Phi^{(s)}_{\text{AA}}$ of associated quantum amplitudes will be 
accumulated during propagation between $\varphi=0$ and $\pi$ that is essentially
an Aharonov-Anandan phase~\cite{ahaanan} for holes confined in the ring. The
latter can be written as the sum of a magnetic-flux-dependent part (the
Aharonov-Bohm~\cite{ahabohm} phase $2\pi\phi/\phi_0$) and a remainder
$\Phi^{(s)}_{\text{G}}$ that depends on the quantum number $s$ distinguishing
dispersion branches: 
\begin{subequations}
\begin{eqnarray}
\Phi^{(s)}_{\text{AA}} &=& 2 \pi \, \frac{\phi}{\phi_0} + \Phi^{(s)}_{\text{G}} \quad , \\
\Phi^{(s)}_{\text{G}} &=& \pi \left( m_{s+} + m_{s-} - 3 s \right) \quad .
\end{eqnarray}
\end{subequations}
The symmetry $m_{s\sigma}=-m_{-s,-\sigma}$ implies $\Phi^{(+)}_{\text{G}} =
-\Phi^{(-)}_{\text{G}}\equiv \Phi_{\text{G}}$. We plot $\Phi_{\text{G}}$ for a realistic
set of parameters in Fig.~\ref{fig:Berry}.

{\it Conclusions and discussion.}\/
We investigated spin-3/2 hole states confined in a quasi-one-dimensional ring. A
number of controlled approximations were employed that can be systematically
improved upon. We find a previously neglected energy-dependent contribution
$\Phi_{\text{G}}$ to the AA phase that results from HH-LH mixing and may be related
to anomalous spin precession~\cite{roland:prl:06} of spin-3/2 particles.
It is likely that this phase is the origin of numerically observed magneto-oscillations of
the conductance polarization in multiply connected hole
nanostructures~\cite{mingwei:pla:06} that persist even in the presence of relatively 
strong disorder.
In a magneto-conductance experiment, the value $\Phi_{\text{G}}(E_{\text{F}})$ of AA
phase for states at the Fermi energy would be observed as a shift in the
Aharonov-Bohm oscillations. This mimicks behaviour
expected~\cite{nitta:apl:99,jairo:prb:07} in systems with a finite zero-field (Rashba or
Dresselhaus) spin splitting, which was used to interpret experimental
data~\cite{mole:prl:06}. We show that, in general, both HH-LH mixing and
spin-splitting effects need to be considered. In addition, the coupling of the ring to
external leads needs to be well-understood, because the character of injected hole
states will depend sensitively on the lead confinement.

{\it Acknowledgment.}\/
UZ thanks P.~Brusheim and D.~Csontos for useful discussions.


\begin{thebibliography}{37}
\bibitem{ApproxComm}
Our approximations amount to neglecting higher-order corrections in $\bar\gamma$
and assuming $\lambda_R\gg |m|$, which is the typical regime of interest. The
condition $\lambda_d<1$ (required for neglecting higher 2D quantum-well bound
states) is usually satisfied.
\bibitem{ProfComm}
The detailed form of spinor components in real space will influcence the coupling of
leads to the ring. However, this is irrelevant for the AA phase, which is our current
focus.
\end{thebibliography}

\end{document}